\documentclass[%
 reprint,
 superscriptaddress,
nofootinbib,
 amsmath,amssymb,
 aps,
]{revtex4-2}

\usepackage{graphicx}
\usepackage{dcolumn}
\usepackage{bm}
\usepackage{hyperref}
\usepackage{braket}
\usepackage{float}

\usepackage{xcolor}
\definecolor{forestgreen}{rgb}{0.13, 0.55, 0.13}

\usepackage[normalem]{ulem}

\begin{document}

\preprint{APS/123-QED}

\title{On the rank-reduced relativistic coupled cluster method}

\author{Alexander V. Oleynichenko}
\email{oleynichenko\_av@pnpi.nrcki.ru (corresponding author)}
\affiliation{
Petersburg Nuclear Physics Institute named by B.P. Konstantinov of National Research Centre ``Kurchatov Institute'', Orlova roshcha 1, Gatchina, Leningradskaya Oblast, 188300, Russia
}
\affiliation{
Moscow Center for Advanced Studies, 20 Kulakova Str., 123592 Moscow, Russia
}

\author{Artem S. Rumiantsev}
\email{rumyantsev\_as@pnpi.nrcki.ru}
\affiliation{
Petersburg Nuclear Physics Institute named by B.P. Konstantinov of National Research Centre ``Kurchatov Institute'', Orlova roshcha 1, Gatchina, Leningradskaya Oblast, 188300, Russia
}
\affiliation{
Saint Petersburg State University, 7/9 Universitetskaya nab., 199034 St. Petersburg, Russia
}

\author{Andr\'{e}i Zaitsevskii}
\email{zaitsevskii\_av@pnpi.nrcki.ru}
\affiliation{
Petersburg Nuclear Physics Institute named by B.P. Konstantinov of National Research Centre ``Kurchatov Institute'', Orlova roshcha 1, Gatchina, Leningradskaya Oblast, 188300, Russia
}
\affiliation{
Department of Chemistry, M. V. Lomonosov Moscow State University, 119991 Moscow, Russia
}

\author{Ephraim Eliav}
\email{ephraim@tauex.tau.ac.il}
\affiliation{
School of Chemistry, Tel Aviv University, 6997801 Tel Aviv, Israel
}

\date{\today}

\begin{abstract}
An efficiency of the Tucker decomposition of amplitude tensors within the single-reference relativistic coupled cluster method with single and double excitations (RCCSD) was studied in a series of benchmark calculations for (AuCl)$_n$ chains, Au$_n$ clusters, and the cluster model of solid YbCl$_2$. The 1~kJ/mol level of accuracy for correlation energy estimates of moderate-size systems and typical reaction energies can be achieved with relatively high compression rates of amplitude tensors \textit{via} rejecting singular values smaller than $\sim 10^{-4}$. For the most extensive system studied (YbCl$_7$ cluster used for modeling of ytterbium center in ytterbium dichloride crystal), only $\sim 3$\% of compressed doubles amplitudes were shown to be significant. Thus, the rank reduction for the relativistic CCSD theory improving its computational scaling is feasible. The advantage (if not necessity) of using the Goldstone diagrammatic technique rather than the ``antisymmetrized'' Brandow one is underlined. The proposed approach is promising for high-precision modeling of relatively large systems with heavy atoms.
\end{abstract}

\keywords{relativistic coupled cluster method; tensor de\-com\-po\-si\-tion; generalized relativistic pseudopotentials; gold clusters; compound-tunable embedding potential}

\maketitle


\section{Introduction}
\label{sec:intro}

Coupled cluster (CC) theory is a versatile and highly accurate tool to predict electronic structure and properties of atoms, molecules, and solids~\cite{Bartlett:07,Zhang:19,Bartlett:24}. Relativistic versions of coupled cluster methods for both closed-shell and open-shell systems were also devised~\cite{Eliav:94,Visscher:96,Visscher:01,Nataraj:10,Eliav:Review:22} and implemented as highly effective computer codes~\cite{Saue:20,Kallay:20,Oleynichenko:EXPT:20,Pototschnig:21}, making it possible to obtain valuable information even on the heaviest elements of the Periodic table, for which the ordinary non-relativistic CC theory gives senseless results.

The common point of different CC theories is their high formal computational cost. Even for the most widely used CCSD model~\cite{Purvis:82} accounting for single and double excitations in the cluster operator, it requires $O(N^6)$ floating-point operations to solve equations for cluster amplitudes, where $N$ stands for the number of one-electron functions (spin-orbitals or spinors
). Memory requirements for the conventional CCSD model scale as $O(N^4)$. These scaling features indicate that the classical algorithms implementing CCSD can be applied to treat only moderate-size molecules, including up to 2-3 dozen atoms, even if the most advanced implementations like that presented in~\cite{Anisimov:14,Kaliman:17} are used. The situation is more severe for the CCSD(T) model~\cite{Raghavachari:89} referred to as the ``gold standard'' of quantum chemistry (with the $O(N^7)$ cost) and for more advanced approximations which fully include contributions from connected triple excitations ($O(N^8)$ for CCSDT~\cite{Noga:87}) and, possibly, quadruples (CCSDTQ-1, etc~\cite{Kucharski:89,Bomble:05}). The computational cost of the relativistic counterparts of these methods scales similarly~\cite{Visscher:96,Visscher:01,Nataraj:10,Oleynichenko:CCSDT:20}, but the prefactor can be 1-2 orders higher due to the lack of spin symmetry and the need to use complex arithmetic for systems with relatively low spatial symmetry~\cite{Saue:99}.

There are two basic strategies to reduce the high computational cost of the coupled cluster theory. The first family of reduced-cost approaches relies on the spatial locality of electron correlation \textit{via} localization of one-electron orbitals~(\cite{Schutz:01,Riplinger:13a,Nagy:17} and references therein). The second possible way is to exploit somehow the low effective rank of tensors representing molecular integrals and cluster amplitudes~\cite{footnote:def-tensors}. This can be achieved by the tensor decomposition techniques (\cite{Kolda:09,Khoromskij:09} and references therein). The basic idea is to represent a high-order tensor as a contraction of two or more tensors with lower orders. Such a decomposition results in a reduction of data amount. It only leads to substantial computational savings if the dimensions of the lower-order tensors (sometimes referred to as effective \textit{tensor ranks}) are comparable to those of the initial tensor. In the specific case of the electronic structure theory, tensor ranks must scale roughly linearly with respect to the system size. In a typical situation effective tensor ranks are defined by the target accuracy requirements.

The most known example of a tensor decomposition with rank reduction in electronic structure theory is the Cholesky decomposition (CD) of the four-index tensor of two-electron (or electron repulsion, ERI) integrals~\cite{Aquilante:11,Pedersen:23},
\begin{equation}
\label{eq:df}
\braket{pq|rs} \approx \sum_P^{N_{\rm DF}} B^P_{pr} B^P_{qs},
\end{equation}
where $B^P_{pr} \equiv (pr|P)$ is the tensor of three-index two-electron integrals and $N_{\rm DF}$ stands for the effective rank of the ERI tensor; in most practical applications, this decomposition is approximated within the density-fitting (DF) technique. The CD and DF approximations are widely used to reduce the computational cost of four-index integral transformation and storage requirements at the correlation calculation stage. Still, they do not allow for reducing the asymptotic computational complexity of CC methods without introducing additional approximations. Many implementations of the non-relativistic DF-CC were reported in the last decades~\cite{Nottoli:23} (and references therein), while the relativistic DF-MP2 and DF-CC methods were presented only recently~\cite{HelmichParis:19,Zhang:24,Chamoli:25}. More aggressive factorizations of the ERI tensor assuming further unpinning of orbital indices in Eq.~(\ref{eq:df}) are also possible, e.~g. the pseudospectral (PS)~\cite{Martinez:PS:95} and chain-of-spheres (COS)~\cite{Dutta:16} approaches, tensor hypercontraction (THC)~\cite{Hohenstein:12a,Hohenstein:12b,Parrish:12,Parrish:13a,Parrish:13b}, canonical product (CP)~\cite{Benedikt:11}, and matrix product state / operator (MPS/MPO)~\cite{Benedikt:MPS:13} decompositions. Generalizing these approaches to CC amplitude tensors is not straightforward since, in contrast to many-body perturbation theories like MP2 and MP3, cluster amplitudes typically cannot be expressed using explicit relations and are obtained in an iterative process. Moreover, the efficiency of applying a particular form of a tensor decomposition to amplitude arrays is questionable and should be firmly established in preliminary numerical tests. One should also avoid decomposing amplitude tensors at each iteration. Despite these difficulties, several schemes for the coupled cluster theory employing decompositions of a tensor of cluster amplitudes were proposed and implemented, e.~g. the CP-based CCD method~\cite{Benedikt:CP:CCD:13}, THC-CC2~\cite{Hohenstein:13a,Hohenstein:13b}, THC-CCSD~\cite{Parrish:14}. Probably the most successful family of such CC methods usually called the rank-reduced (RR) ones, is based on the approximate Tucker decomposition~\cite{Tucker:66} of CC amplitude tensors. Within this approach, cluster amplitudes (e.~g. doubles $t_{ij}^{ab}$) which are nearly linear dependent or close to zero are eliminated by ``compression'' of an amplitude tensor according to
\begin{equation}
\label{eq:non-rel-tucker-ccsd}
t_{ij}^{ab} \equiv t_{ia,jb} \approx \sum_{XY}^{N_{\rm SVD}} T_{XY} U^X_{ia} U^Y_{jb},
\end{equation}
where $N_{\rm SVD}$ stands for the effective tensor rank of $t_{ia,jb}$ and the $T_{XY}$ matrix contains compressed cluster amplitudes~\cite{Parrish:19} (similar expressions can also be introduced for higher excitations~\cite{Lesiuk:CC3:19,Lesiuk:CCSDTQ:22}). The $U_{ia}^X$ projectors are obtained before the start of CC iterations by an eigendecomposition of an approximate amplitude tensor obtained within the MP2 or MP3 methods~\cite{Parrish:19} (or higher-order singular value decomposition for CC models accounting for triple excitations). By setting the threshold $\varepsilon$ of retained eigenvalues (or, more general, singular values), one can control the target accuracy of the representation~(\ref{eq:non-rel-tucker-ccsd}). The high computational efficiency of the rank-reduced coupled cluster method is possible since for the given threshold $\varepsilon$ the effective tensor rank $N_{\rm SVD}$ scales roughly linearly with respect to the system size~\cite{Parrish:19,Lesiuk:CC3:19}; for relatively large systems $N_{\rm SVD} \ll N_{\rm occ} N_{\rm virt}$ ($N_{\rm occ}$ and $ N_{\rm virt}$ stand for the number of occupied and virtual one-electron functions, respectively). Combined with the density fitting approximation~(\ref{eq:df}) (and under some additional controllable approximations) the RR-CCSD method was shown to scale as $O(N^5)$~\cite{Lesiuk:22} or even $O(N^4)$ if the THC decomposition is applied to both projectors $U_{ia}^X$ and two-electron integrals~\cite{Hohenstein:22}. In the recent few years the rank-reduced versions of the most practically important CC methods beyond CCSD were also devised and implemented, including CCSD(T)~\cite{Lesiuk:22,Zhao:24}, CCSDT-1~\cite{Hino:04}, CC3~\cite{Lesiuk:CC3:19}, CCSDT~\cite{Lesiuk:CCSDT:19,Lesiuk:21}, CCSDT(Q)~\cite{Lesiuk:CCSDTQ:22}, EOM-CCSD~\cite{Hohenstein:19} and EOM-CC3~\cite{Michalak:24}.

All the cited works considered only the non-relativistic version of the coupled cluster theory for closed-shell systems; the derived equations were formulated from the very beginning in the spin-adapted formalism. From the practical point of view, the relativistic version of the rank-reduced CC method is desired. Moreover, it is natural to work with the Kramers-unrestricted version of RCC~\cite{Visscher:96} since it is pretty general and provides a possibility to treat ``high-spin'' open shell systems very typical for heavy-element chemistry and physics (for example, ground states of most lanthanide and actinide compounds are of this type). As seen later, the generalization of the RR-CC method to the relativistic case is relatively straightforward, but several points must be carefully addressed. These issues are considered in detail in the present work.

The paper is organized as follows. In Section~\ref{sec:theory}, we briefly recapitulate the basics of the single-reference CC method, introduce a generalization of the Tucker decomposition~(\ref{eq:non-rel-tucker-ccsd}) for the relativistic case, and formulate working equations of the relativistic RR-CC method. Sections~\ref{sec:comp} and~\ref{sec:results} focus on the singular structure of RCCSD amplitude tensors for model systems and study the convergence of correlation energy and relative energies with respect to the system size and the singular value threshold. Section~\ref{sec:comp} introduces three model systems, i.~e. (AuCl)$_n$ chains, Au$_n$ clusters and the cluster model of solid YbCl$_2$, while the results obtained for these models are further discussed in details in Section~\ref{sec:results}. Section~\ref{sec:conc} provides conclusive remarks and briefly outlines the prospects.

We use the following convention for labeling of different classes of indices:
\begin{itemize}
\item $p, q, r, s, ...$: general index for molecular spinors, range is $[1,N]$;
\item $i, j, k, l, ...$: occupied molecular spinors, range is $[1,N_{\text{occ}}]$;
\item $a, b, c, d, ...$: virtual molecular spinors, range is $[1,N_{\text{virt}}]$;
\item $P, Q, ...$: auxiliary DF/CD basis functions, range is $[1,N_{\text{DF}}]$;
\item $X, Y, ...$: auxiliary index for compressed cluster amplitudes, range is $[1,N_{\text{SVD}}]$.
\end{itemize}

\section{Theoretical background}
\label{sec:theory}

\subsection{Single-reference relativistic coupled cluster method}

Within the single-reference coupled cluster method, an electronic wave function $\ket{\psi}$ is obtained by acting on the vacuum (typically Hartree-Fock) determinant $\ket{\Phi_0}$ by the exponentially parameterized wave operator:
\begin{equation}
\label{eq:cc-ansatz}
\ket{\psi} = e^T \ket{\Phi_0},
\end{equation}
where $T$ stands for the cluster operator. For the CCSD approximation considered in the present work it consists of only single and double excitation operators, $T = T_1 + T_2$, where $T_1 = \sum\limits_{ia} t_i^a a_a^\dagger a_i$. There are two possible ways to define the $T_2$ operator~\cite{footnote:def-t2}. Within the first one, it is parameterized as~\cite{Lindgren:78,Matthews:19}
\begin{equation}
\label{eq:def-t2-goldstone}
T_2 = \frac{1}{2} \sum_{ijab} t_{ij}^{ab} a_a^\dagger a_b^\dagger a_j a_i,
\end{equation}
where the tensor of cluster amplitudes $t_{ij}^{ab}$ is symmetric with respect to the simultaneous permutation of electrons, $t_{ij}^{ab} = t_{ji}^{ba}$. This version of the CC formalism uses Goldstone diagrams to derive working equations. It is used mainly in atomic CC programs (since it allows for the convenient separation of angular variables). In contrast, most of modern molecular programs implementing RCC are based on a more compact antisymmetrized formalism of Brandow diagrams, which uses a slightly different form of the $T_2$ operator~\cite{ShavittBartlett:09}:
\begin{equation}
\label{eq:def-t2-brandow}
T_2 = \frac{1}{4} \sum_{ijab} \mathcal{T}_{ij}^{ab} a_a^\dagger a_b^\dagger a_j a_i,
\end{equation}
where the amplitude tensor $\mathcal{T}_{ij}^{ab} = t_{ij}^{ab} - t_{ij}^{ba}$ is antisymmetric with respect to the $i\leftrightarrow j$ or $a\leftrightarrow b$ permutations:
\begin{equation}
\label{eq:antisym}
\mathcal{T}_{ij}^{ab} =
- \mathcal{T}_{ji}^{ab} =
- \mathcal{T}_{ij}^{ba} =
\mathcal{T}_{ji}^{ba}
\end{equation}
The decomposition~(\ref{eq:non-rel-tucker-ccsd}) does not obey this antisymmetry property and thus cannot be directly applied to the $\mathcal{T}_{ij}^{ab}$ tensor. This problem does not arise for the $t_{ij}^{ab}$ amplitudes defined by Eq.~(\ref{eq:def-t2-goldstone}). For this reason, the relativistic RR-CC theory must be formulated in terms of the Goldstone formalism despite the inevitable (but still manageable) growth in the number of diagrams for CC models accounting for higher excitations beyond doubles.

The working equations for the cluster amplitudes $t_i^a$ and $t_{ij}^{ab}$ are obtained by inserting the Ansatz~(\ref{eq:cc-ansatz}) into the two- or four-component relativistic counterpart of the electronic Schr\"{o}dinger equation $H\ket{\psi} = E\ket{\psi}$ and projecting it onto the manifold of singly ($\Phi_i^a$) and doubly ($\Phi_{ij}^{ab}$) excited determinants:

\begin{equation}
\label{eq:singles-eq}
t_i^a \varepsilon_i^a = - \braket{ \Phi_i^a | (V e^T)_c | \Phi_0 } = - r_{i}^{a},
\end{equation}
\begin{equation}
\label{eq:doubles-eq}
t_{ij}^{ab} \varepsilon_{ij}^{ab} = - \braket{ \Phi_{ij}^{ab} | (V e^T)_c | \Phi_0 } = - r_{ij}^{ab},
\end{equation}
where $\varepsilon_i^a = \varepsilon_a - \varepsilon_i$ and $\varepsilon_{ij}^{ab} = \varepsilon_i^a + \varepsilon_j^b = \varepsilon_a + \varepsilon_b - \varepsilon_i - \varepsilon_j$ are energy denominators and the $c$ subscript means that only terms represented by connected diagrams are retained. $V$ stands for the difference between the mean-field and true many-electron Hamiltonians. The number of floating-point operations required to solve Eqs.~(\ref{eq:singles-eq}) and~(\ref{eq:doubles-eq}) scales as $O(N^2_{occ}N^4_{virt})$, while the memory requirements needed to store doubles amplitudes scale as $O(N^2_{occ}N^2_{virt})$.

\subsection{Tucker decomposition of the amplitude tensor and obtaining projectors}

Within the non-relativistic reduced-rank coupled cluster method the computational cost of Eqs.~(\ref{eq:singles-eq}) and~(\ref{eq:doubles-eq}) is lowered by a simultaneous application of the density fitting approximation (Eq.~(\ref{eq:df})) and the Tucker decomposition of amplitudes $t_{ij}^{ab}$ (singles remain uncompressed)~\cite{Parrish:19,Lesiuk:22,Hohenstein:22}. High computational efficiency and accuracy of the non-relativistic RR-CC methods are mainly determined by two factors, (a) nearly optimal projectors $U_{ai}^X$ are obtained by the eigendecomposition of MP3 amplitudes, and (b) the tensor rank $N_{\rm SVD}$ scales approximately linearly with respect to the system size. The fulfillment of the latter issue was examined in a series of numerical tests (see below in Section~\ref{sec:results}). Here, we focus on possible generalizations of the decomposition~(\ref{eq:non-rel-tucker-ccsd}) for the relativistic case.

For both closed-shell systems with low spatial symmetry and open-shell systems treated within the Kramers-unrestricted approach, all tensors arising in the relativistic coupled cluster method, including molecular integrals and amplitudes, are complex-valued~\cite{Saue:99}. Thus, the supermatrix $t_{ia,jb}$ representing an amplitude tensor is a complex symmetric matrix. The existence of eigenvalues for such matrices is not guaranteed (in contrast to the non-relativistic case where the symmetric $t_{ia,jb}$ matrix is real and thus Hermitian). One should obtain Tucker projectors not by an eigendecomposition but in another way. The most obvious solution is to use the singular value decomposition (SVD) of a matrix $\bm{t}$ of MP2 or MP3 amplitudes,
\begin{equation}
\bm{t} = \bm{U} \bm{\Sigma} \bm{V}^\dagger,
\end{equation}
where both $\bm{U}$ and $\bm{V}$ are unitary matrices, and $\bm{\Sigma}$ is a diagonal matrix of singular values. Due to the symmetry of the matrix of cluster amplitudes in the relativistic case the ansatz (Eq.~(\ref{eq:non-rel-tucker-ccsd})) remains the same, but the projectors become complex:
\begin{equation}
\label{eq:rel-tucker-ccsd}
t_{ia,jb} \approx \sum_{XY} T_{XY} U^X_{ia} U^{Y}_{jb}.
\end{equation}
One of the most important advantages of the SVD decomposition is the existence of highly efficient algorithms allowing to obtain only singular values above the pre-defined threshold, e.~g. the partial Golub-Kahan algorithm~\cite{Golub:65,Simon:00,Baglama:05}. If both the DF approximation and the Laplace transformation of energy denominators~\cite{Haser:92} are used, its computational cost can be reduced to $O(N^4)$ and $O(N^5)$ for the MP2 and MP3 projectors, respectively~\cite{Lesiuk:22}. The latter can be further reduced to $O(N^4)$ by employing the THC technique~\cite{Hohenstein:12a}. Finally, SVD is also applicable to non-square complex matrices, which would be of crucial importance for the forthcoming adaptation of the rank-reduced approach to the relativistic Fock-space multireference coupled cluster method~\cite{Visscher:01,Eliav:Review:22}.

The other possibility to obtain Tucker projectors to be mentioned is the Takagi decomposition of a complex symmetric matrix~\cite{Takagi:90,Dieci:22},
\begin{equation}
\bm{t} = \bm{U} \bm{\Sigma} \bm{U}^T,
\end{equation}
where $\bm{U}$ is a unitary matrix. This decomposition allows for using the non-relativistic formula (\ref{eq:non-rel-tucker-ccsd}). Both the Takagi and SVD decompositions bear the same set of singular values. Despite efficient algorithms for the Takagi factorization being reported~\cite{Xu:08,Chebotarev:14}, its high-performance program implementations are still not widely available, hampering its usefulness for the relativistic RR-CC method. Thus, it is not used in the present work.

\subsection{Working equations for compressed cluster amplitudes}
\label{sec:work-eqs}

One of the most attractive features of the rank-reduced CC theory is the possibility to reformulate its working equations entirely in terms of compressed amplitudes $T_{XY}$. The costly singular value decomposition is performed only once before starting CC iterations. To obtain the explicit equations for $T_{XY}$ one should multiply both sides of Eq.~(\ref{eq:doubles-eq}) by $U^{X'}_{ia} U^{Y'*}_{jb}$ and sum over the $i,j,a,b$ indices taking into account the unitarity of projectors
\begin{equation}
\label{eq:unitary}
\sum_{ia} U^{X}_{ia} U^{X'*}_{ia} = \delta_{XX'}.
\end{equation}
One arrives at the matrix equations defining compressed amplitudes:
\begin{equation}
\label{eq:compressed-matrix-eq}
\bm{E}\bm{T} + \bm{T}\bm{E} = - \bm{R},
\end{equation}
where $R_{X'Y'} = \sum\limits_{X'Y'} r_{ij}^{ab} U^{X'*}_{ia} U^{Y'*}_{jb}$ stands for the transformed residual matrix and the transformed energy denominator matrix $\bm{E}$ is defined as
\begin{equation}
\label{eq:def-E-matrix}
E_{XX'} = \sum_{ia} \varepsilon_i^a\ U^{X*}_{ia} U^{X'}_{ia}.
\end{equation}
Eq.~(\ref{eq:compressed-matrix-eq}) coincides with its nonrelativistic counterpart (Eq.~(38) in Ref.~\cite{Parrish:19}). Since the $\bm{E}$ matrix is Hermitian, one can diagonalize it
\begin{equation}
\bm{E} = \bm{W} \tilde{\bm{\varepsilon}} \bm{W}^\dagger,
\end{equation}
and then simplify the Eq.~(\ref{eq:compressed-matrix-eq}):
\begin{equation}
\label{eq:compressed-eq}
(\tilde{\varepsilon}_X + \tilde{\varepsilon}_Y) \tilde{T}_{XY} = - \tilde{R}_{XY},
\end{equation}
where $\tilde{\bm{T}} = \bm{W}^\dagger \bm{T} \bm{W}$ and $\tilde{\bm{R}} = \bm{W}^\dagger \bm{R} \bm{W}$. Equations~(\ref{eq:compressed-eq}) are then solved with respect to $\tilde{T}_{XY}$.

In principle one can adapt the factorization~(\ref{eq:rel-tucker-ccsd}) to represent antisymmetrized cluster amplitudes defined by Eq.~(\ref{eq:antisym}):
\begin{equation}
\mathcal{T}_{ia,jb} = \sum_{XY} T_{XY} \left( U_{ia}^X U_{jb}^{Y} -  U_{ib}^X U_{ja}^{Y} \right)
\end{equation}
(cf. Eq.~(29) in~\cite{Chamoli:25} for the antisymmetrized tensor of two-electron integrals in the relativistic DF-CC method). However, in this case we have not found a way to reformulate CC amplitude equations~(\ref{eq:doubles-eq}) in terms of $T_{XY}$ in a computationally efficient manner.

Working expressions for the right-hand side of the relativistic RR-CC equations (the transformed residual matrix $R_{XY}$) are obtained straightforwardly by the projection of the right-hand side of Eqs.~(\ref{eq:singles-eq}) and~(\ref{eq:doubles-eq}) onto the ``subspace'' of compressed amplitudes. It is necessary to take care of the correct order of tensor contractions to ensure the least possible asymptotic complexity. The complexity of the most computationally demanding (the so called ``ladder'') term $\sum\limits_{cd} \braket{ab|cd} t_{ij}^{cd}$ within the reduced-rank approximation is reduced from $O(N_{occ}^2 N_{virt}^4)$ to $O(N_{occ} N_{virt} N_{DF} N_{SVD}^2) \sim O(N^5)$ in a complete analogy with the non-relativistic case~\cite{Parrish:19}. The only problem is the presence of the $O(N^6)$-scaling terms that are quadratic in amplitudes $t_{ij}^{ab}$, as their computational complexity cannot be reduced in this way. The possible solution proposed in~\cite{Lesiuk:22} involves the introduction of additional Tucker decompositions of intermediate tensors $O_{ki,lj} = \sum\limits_{cd} \braket{kl|cd} t_{ij}^{cd}$ and $Z_{jb,kc} = \sum\limits_{ld} \braket{kl|dc} t_{jl}^{bd}$, which results in the further reduction of the overall scaling to $O(N^5)$ (see Appendix~\ref{sec:appendix} for the detailed discussion). The complete list of terms arising in the relativistic RR-CCD approximation (double excitations only), together with estimates of their scaling, is given in Supplementary Materials.

\section{Model systems and computational details}
\label{sec:comp}

\begin{figure}
    \centering
    \includegraphics[width=\linewidth]{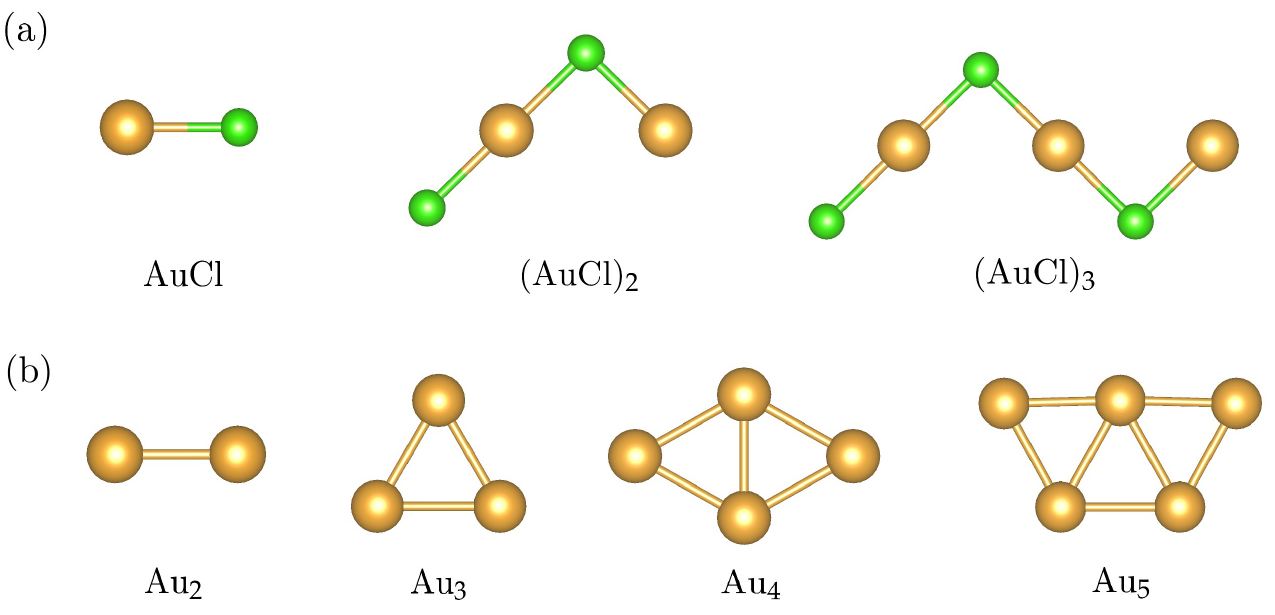}
    \caption{Molecular structures of: (a) (AuCl)$_n$ chains ($n=1-3$), (b) Au$_n$ clusters ($n=2-5$). Au and Cl atoms are shown in gold and green colours, respectively.}
    \label{fig:au-clusters}
\end{figure}

\begin{figure}
    \centering
    \includegraphics[width=0.9\linewidth]{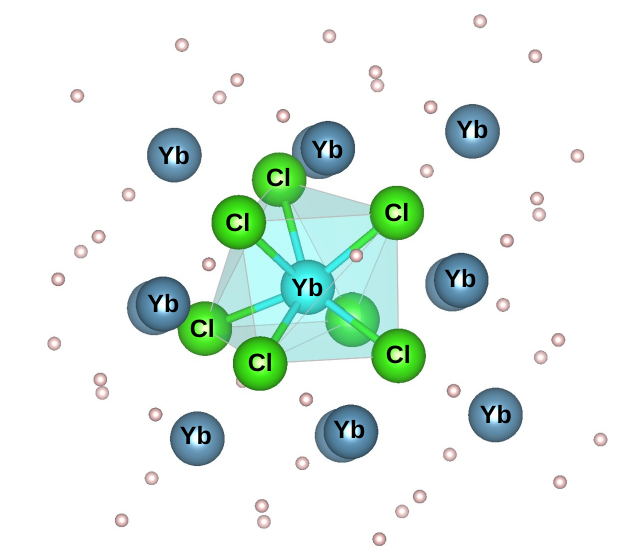}
    \caption{The structure of the cluster model of solid 
ytterbium(II) chloride~\cite{Shakhova:22} (see the text for details). Chemical bonds and the coordination polyhedron are shown for the main cluster. The central Yb atom, Cl atoms, and Yb pseudoatoms are shown in cyan, green, and blue, respectively. The chlorine pseudoatoms of the third coordination sphere, replaced by fractional point charges, are shown in faded red.}
    \label{fig:ybcl2-cluster}
\end{figure}

We performed a series of benchmark calculations to elucidate the efficiency of representing the cluster operator $T_2$ (Eq.~(\ref{eq:def-t2-goldstone})) using the truncated Tucker decomposition and test the accuracy of this approximation at different singular value thresholds. In most other works dealing with the non-relativistic closed-shell coupled cluster theory, the typical objects used in such calculations were molecules consisting of the 1st and 2nd-row element atoms (\cite{Lesiuk:CC3:19,Parrish:19,Lesiuk:22} and references therein). These objects are irrelevant for assessing the features of the relativistic coupled cluster theory. Thus, in the present work, we propose three types of molecular systems whose electronic structure clearly cannot be modelled in the non-relativistic approximation: (a) gold(I) monochloride chains (AuCl)$_n$, (b) gold clusters Au$_n$, and (c) the cluster model of the 
ytterbium(II) chloride
crystal (see Figs.~\ref{fig:au-clusters},\ref{fig:ybcl2-cluster}).

The (AuCl)$_n$ chains form gold(I) monochloride crystals~\cite{Janssen:74}. From the conceptual point of view, these model systems can be considered as relativistic counterparts of effectively one-dimensional molecules such as polyenes or alkanes. Due to the technical limitations, here we consider only the AuCl, (AuCl)$_2$, and (AuCl)$_3$ molecules (Fig.~\ref{fig:au-clusters}a). These limitations can be lifted by a more elaborate treatment of the Coulomb tensor at the transformation stage (this work is currently in progress and will be published elsewhere). Bond length $r$(AuCl) = 2.3559~\AA\ and valence angles $\angle$(Cl-Au-Cl) = 180.0$^\circ$, $\angle$(Au-Cl-Au) = 91.2$^\circ$ were taken from the crystallographic data for solid AuCl~\cite{Janssen:74}. These molecules are closed-shell systems.

\begin{figure}
\centering
\includegraphics[width=\linewidth]{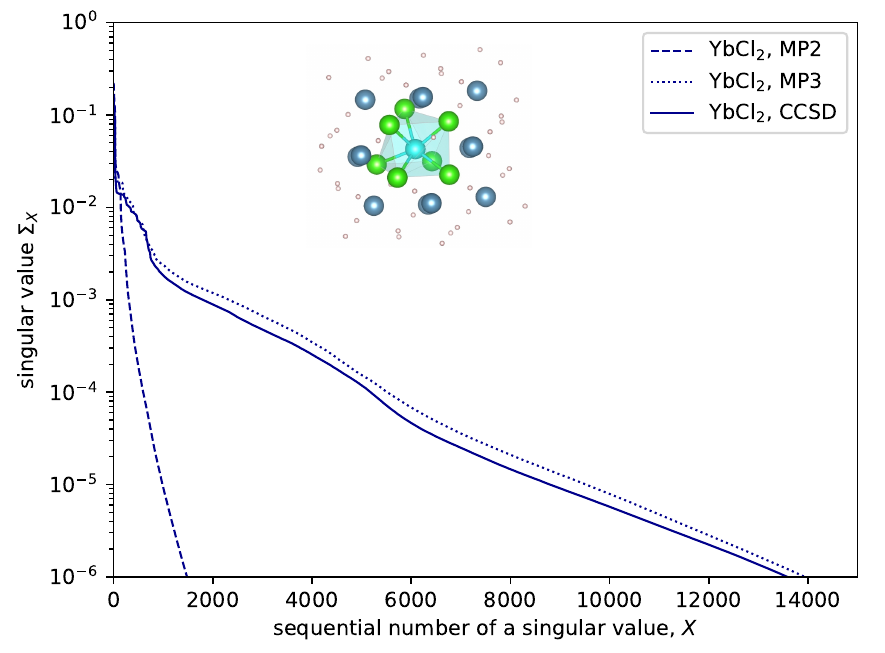}
\caption{Singular value distribution for the $t_{ij}^{ab}$ amplitude tensor for the YbCl$_7$@CTEP cluster. $\Sigma_X$ denotes singular values (cf.~Eq.~(9)). Singular values are sorted in the descending order.
\label{fig:singular-values}}
\end{figure}

The second class of molecular systems considered are small gold clusters Au$_n$, $n=1-5$. These systems were intensively studied theoretically and experimentally in the last two decades due to their pronounced catalytic activity~\cite{Xiao:06,Assadollahzadeh:09,Takei:12} (and references therein). Gold clusters were also used as model systems in theoretical studies of adsorption of short-lived superheavy element atoms on a gold surface in thermochromatographic experiments~\cite{Pershina:15,Rusakov:13,Demidov:24}. Most theoretical studies of gold clusters published to date employed the scalar-relativistic density functional (DFT) approach. The notable exception is the Au$_3$ cluster thoroughly studied using the relativistic DFT and CCSD(T) methods, which predict the equilateral triangle geometry for it~\cite{Rusakov:07,Rusakov:23} (in contrast to all other approaches not accounting for the spin-orbit coupling). The scalar-relativistic CCSD(T) calculations were also reported for the Au$_4$ -- Au$_8$ clusters~\cite{Baek:17}. Thus the study of small gold clusters by the fully relativistic CC method is very promising for resolving the remaining controversial issues in this field. Here we focus on the Au$_1$ -- Au$_5$ clusters (see Fig.~\ref{fig:au-clusters}b). For Au$_2$ we assumed the experimental bond length 2.472~\AA~\cite{Bishea:91}, while the theoretically predicted geometries were used for Au$_3$~\cite{Rusakov:23} and Au$_4$, Au$_5$~\cite{Baek:17}. Clusters containing odd numbers of Au atoms possess open-shell ground states, which can be (very roughly) thought about as spin doublets and were modeled within the Kramers unrestricted HF and CC approaches.

\begin{figure}
\centering
\includegraphics[width=\linewidth]{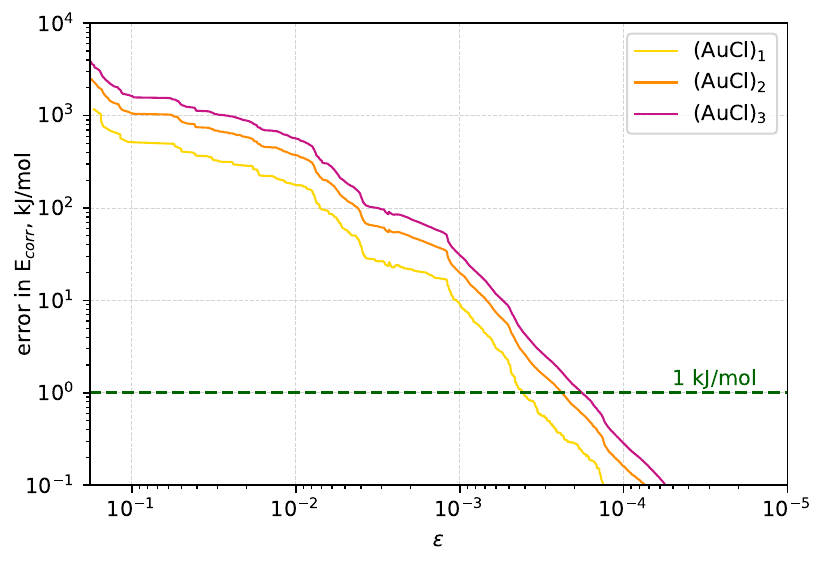}
\includegraphics[width=\linewidth]{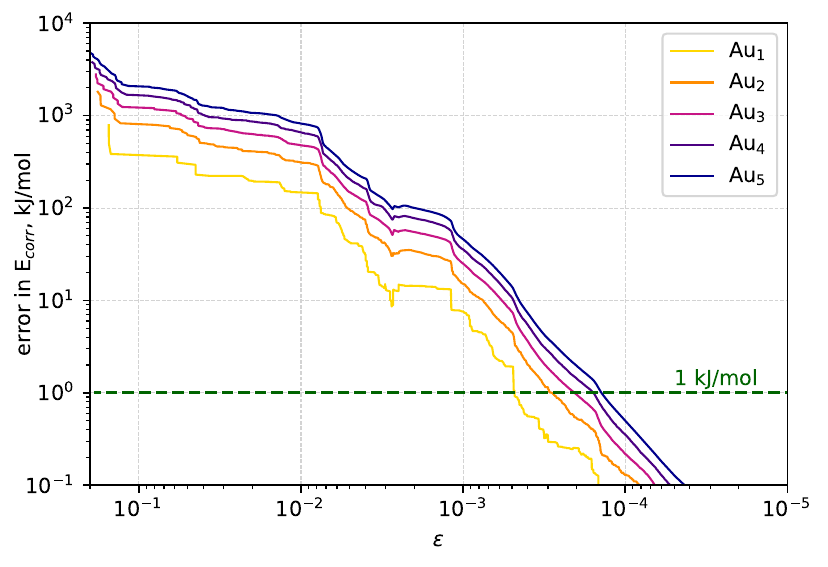}
\includegraphics[width=\linewidth]{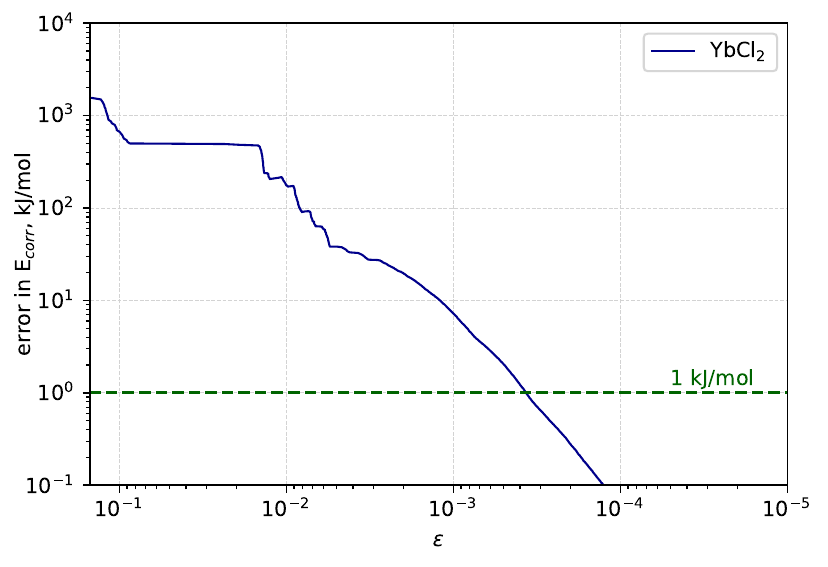}
\caption{Convergence of the CCSD correlation energy (kJ/mol) as a function of the singular value threshold $\varepsilon$ used for compression of cluster amplitudes for (AuCl)$_n$ chains (top panel), Au$_n$ clusters (middle panel), YbCl$_7$@CTEP cluster model (bottom panel). The horizontal green dashed line marks the 1~kJ/mol accuracy level.
\label{fig:ecorr-err}}
\end{figure}

\begin{figure}
\centering
\includegraphics[width=\linewidth]{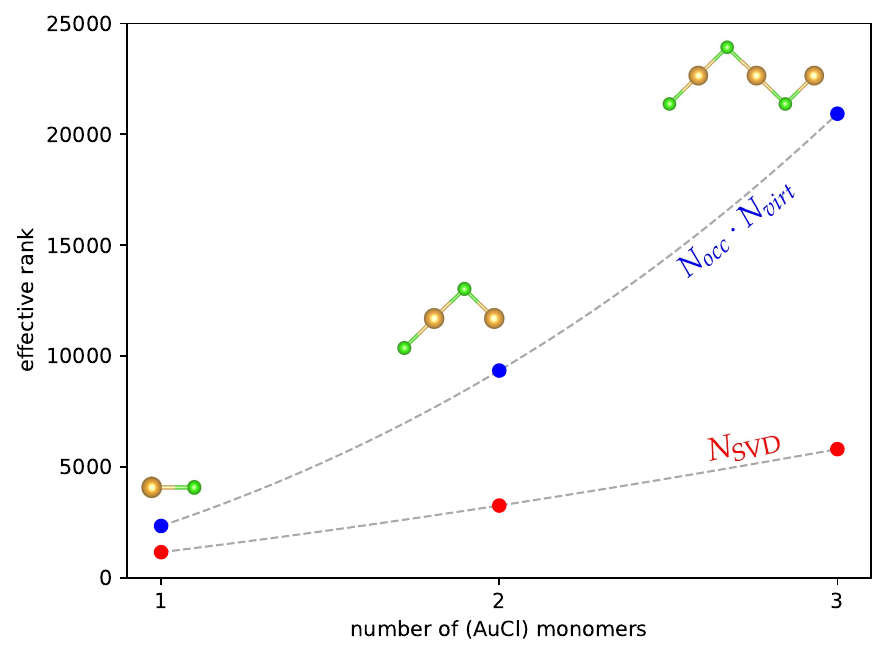}
\includegraphics[width=\linewidth]{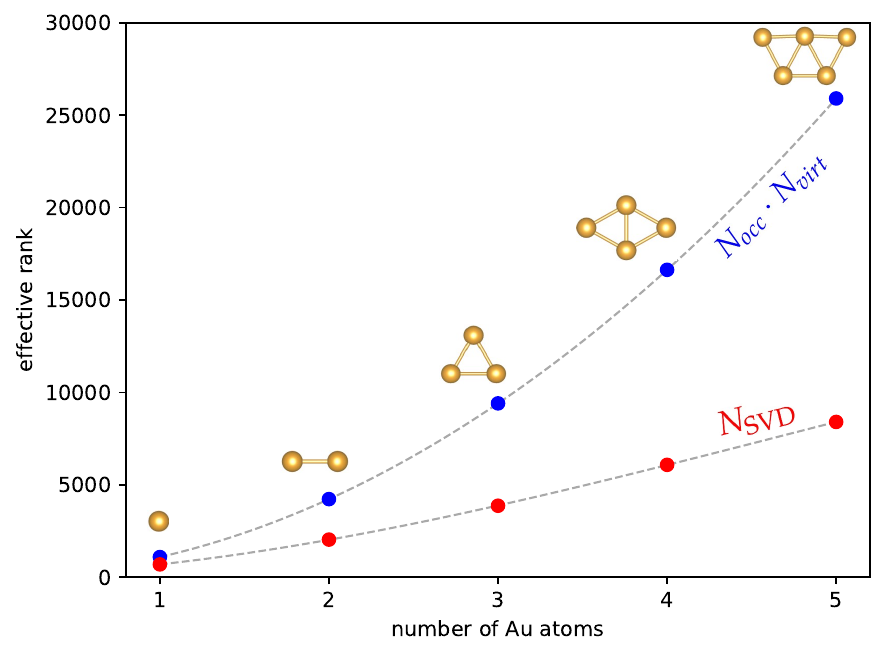}
\caption{Effective rank $N_{\rm SVD}$ of the $t_{ij}^{ab}$ amplitude tensor needed for the 1~kJ/mol accuracy level of the CCSD correlation energy restoration for (AuCl)$_n$ chains (top panel), Au$_n$ clusters (bottom panel). The full rank of a tensor equals $N_{occ}N_{virt}$.
\label{fig:eff-rank}}
\end{figure}

In calculations of both the (AuCl)$_n$ and Au$_n$ systems, relativistic effects were simulated by using the shape-consistent generalized relativistic pseudopotentials of small atomic cores (GRPPs), which effectively include spin-orbit and Breit interactions~\cite{Titov:99,Petrov:04,Mosyagin:10,Oleynichenko:LIBGRPP:23}. The GRPPs employed for Au and Cl exclude 60 and 2 core electrons, respectively, leaving outercore and valence electrons for the explicit treatment~\cite{GRPPs}. At the same time, only $5d6s$ shells of Au and $3s3p$ shells of Cl were further correlated at the CCSD level. Contracted Gaussian basis set $[5s6p5d2f1g]$ for Au was obtained using the relativistic counterpart of the averaged atomic natural orbitals (ANO) procedure~\cite{Almlof:91,Oleynichenko:ANO:24} by averaging relativistic CCSD density matrices obtained for the ground states of the neutral Au atom and the Au$^+$ ion; exponential parameters of primitive Gaussians were taken from the dyall.v4z basis set~\cite{Dyall:04}. Basis set $[4s5p2d1f]$ used for Cl is the straightforward adaptation of the cc-pVTZ set~\cite{Woon:93} for use with the small-core GRPP. Numbers of occupied and virtual spinors used at the CCSD stage for both Au$_n$ and (AuCl)$_n$ systems are given in the Supplementary Materials.

To assess the performance of the Tucker decomposition for extended three-dimensional objects we considered the cluster model of the solid ytterbium (II) chloride~\cite{Shakhova:22}. Ytterbium halides were proposed in~\cite{Shakhova:22,Khadeeva:24} as prototype objects for testing new approaches to calculating local properties depending on electron and spin densities in the vicinity of a heavy nucleus, e.~g. chemical shifts of X-ray emission lines, in $f$-element containing compounds and materials~\cite{Titov:14}. A cluster model of solid constructed within the compound-tunable embedding potential (CTEP) approach~\cite{Maltsev:21,Shakhova:22,Oleynichenko:YPO4:24} assumes that a heavy atom of interest and its first coordination sphere (a main cluster) is treated at the highest possible level of theory. Atoms of the next one or two coordination spheres are approximated by pseudopotentials of a special type with parameters tuned specifically for a given compound. The influence of the remaining part of a crystal is simulated by the set of optimized fractional point charges placed on atoms outside a main cluster.
The cluster model chosen for the benchmark included one central Yb$^{2+}$ ion, 7 Cl$^-$ ions in the first coordination sphere (implying that YbCl$_7^{5-}$ is the main cluster), 12 pseudoions (i.~e. centers with semilocal pseudopotentials implying the exclusion of all electrons of these ions from the explicit treatment) representing Yb$^{2+}$ of the second coordination sphere, and 58 fractional point charges simulating the effect of a crystal (see Fig.~2). This model will be further referred to as YbCl$_{7}$@CTEP.
Contracted Gaussian basis sets for Yb and Cl were of size $[4s4p2d2f]$ and $[5s4p1d]$, respectively. The shape-consistent GRPP replacing 28 core electrons~\cite{GRPPs,Mosyagin:17} was used for the central Yb atom. Pseudoatoms representing the Yb$^{2+}$ ions of the second coordination sphere were supplied with the $[3s2p2d]$ basis sets to allow for a flexible description of the electronic wavefunction at the boundaries of the YbCl$_7$ fragment. At the relativistic CCSD stage 42 occupied and 512 virtual molecular spinors (554 spinors overall) were involved in the correlation treatment.

Kramers-unrestricted Hartree-Fock calculations as well as molecular integral trans\-for\-ma\-tion were performed using the PySCF program package~\cite{Sun:17,Sun:20} extended with the LIBGRPP library to calculate integrals over generalized relativistic pseudopotentials~\cite{Oleynichenko:LIBGRPP:23,Oleynichenko:LIBGRPP:25}. A new implementation of the relativistic CCSD method based on the Goldstone formalism was developed and included into the EXP-T program package (version 1.8.5)~\cite{Oleynichenko:EXPT:20,EXPT:24}. We also developed a new program CCRS oriented specifically at high-performance rank-reduced relativistic coupled cluster calculations of low-symmetric systems. CCRS is written in the Rust programming language; it also provides C and Python interfaces for accessing its subroutines from third-party codes. Molecular structures were visualized using the VESTA program~\cite{Momma:08}. All molecular geometries and parameters of the basis sets used are provided in the Supplementary Materials.

\section{Results and discussion}
\label{sec:results}

We now discuss the efficiency of an approximate Tucker decomposition of cluster amplitude tensors. Fig.~\ref{fig:singular-values} shows the distributions of singular values of the $t_{ij}^{ab}$ tensor for the YbCl$_7$@CTEP cluster model (distributions for (AuCl)$_n$ and Au$_n$ are essentially similar, see Supplementary Materials). It can be seen that the distribution for MP2 amplitudes differs sharply from those for MP3 and converged CCSD amplitudes. A similar picture was previously reported for the non-relativistic case~\cite{Parrish:19} and explained by the fact that the MP2 amplitude matrix is negative-definite. In contrast, for MP3 and CCSD, this matrix is indefinite, resulting in fundamental differences in the spectra compared to MP2. In the relativistic case we use SVD instead of eigendecomposition, but the general trends are still similar. The obtained distributions justify the use of MP3 projectors in the Tucker decomposition (\ref{eq:rel-tucker-ccsd}) to make its length $N_{\rm SVD}$ acceptable while preserving computational gains of its use.

The second point to be addressed is the justified choice of appropriate $N_{\rm SVD}$. For the exact expansion (\ref{eq:rel-tucker-ccsd}) it equals to $N_{occ}N_{virt}$. To achieve the reduced scaling in CC amplitude equations, $N_{\rm SVD}$ should possess lower scaling with respect to the system size; ideally, it should be linear. Restricting the expansion (\ref{eq:rel-tucker-ccsd}) to first $N_{\rm SVD}$ terms one inevitably introduces an error in calculated correlation energy. Thus, $N_{\rm SVD}$ should be defined by the desired level of accuracy. In some previous works, such a level was determined as a relative error in correlation energy; typically, the 0.1\%  error was considered acceptable. However, correlation energy is an extensive property and it scales linearly with respect to the system size, thus such a criterion can also lead to an unacceptable drop in accuracy for large objects. Here, we propose using the absolute error in correlation energy as a target to choose the threshold $\varepsilon$ for singular values (and thus $N_{\rm SVD}$) properly. This target accuracy depends on the field of application and the object considered.
To make the method relevant for most nowadays chemical applications, it is natural to set the required accuracy level to 1~kJ/mol ($\approx$~84~cm$^{-1}$).
Convergence of the CCSD correlation energies $E_{corr}$ for the systems considered with respect to the singular value threshold $\varepsilon$ is shown in Fig.~\ref{fig:ecorr-err}.
Reference values of $E_{corr}$ were obtained by the exact relativistic CCSD method. For the test purposes, converged cluster amplitudes were decomposed, singular values below the given threshold $\varepsilon$ were discarded, and then the full tensor was reconstructed to simulate the rank-reduced CC approach; correlation energies were recalculated for each $\varepsilon$.
The effective tensor ranks $N_{\rm SVD}$ corresponding to the 1~kJ/mol accuracy of correlation energies are plotted in Fig.~\ref{fig:eff-rank}. One can see that $N_{\rm SVD}$ grows slower than the full rank $N_{occ} N_{virt}$. However, the linear scaling is still unreachable for the systems considered here since both the (AuCl)$_n$ chains and especially Au$_n$ clusters are relatively spatially compact objects (the actual scaling is roughly $N_{\rm SVD} \sim N_{occ}^{1.5}$). This result shows that reducing the overall scaling of the relativistic RR-CCSD method is possible. Using the obtained $N_{\rm SVD}$ values, one can estimate the fraction of cluster amplitudes needed to achieve the 1~kJ/mol accuracy level as $N_{\rm SVD}^2 / (N_{occ}N_{virt})^2$. For the largest molecular systems considered here, only 7.7\% for (AuCl)$_3$, 10.5\% (Au$_n$), and 2.6\% (YbCl$_7$@CTEP) of doubles amplitudes are significant. The best compression rate is achieved for the most extended three-dimensional object (YbCl$_7$@CTEP).

\begin{figure}
    \centering
    \includegraphics[width=\linewidth]{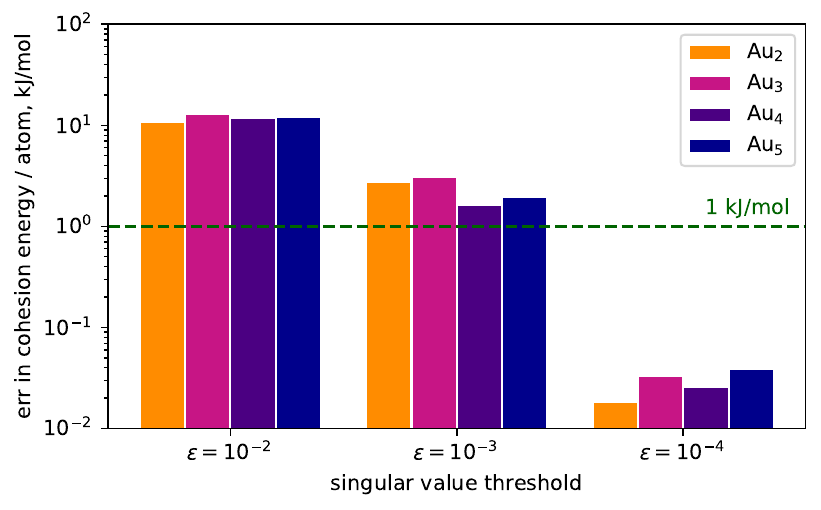}
    \caption{Errors in cohesion energies of Au$_n$ clusters (kJ/mol, per 1 atom) with respect to the singular value threshold $\varepsilon$. The horizontal green dashed line marks the 1~kJ/mol accuracy level.}
    \label{fig:cohesion}
\end{figure}

In almost all practical applications, relative energies rather than absolute ones are of primary interest. In contrast to full correlation energies in the case of relative energies, one can hope for some error compensation, resulting in the possibility of using a less stringent singular value threshold $\varepsilon$. Such a compensation was indeed observed in non-relativistic closed-shell RR-CC calculations~\cite{Lesiuk:22,Hohenstein:22}. We calculated cohesion energies of the Au$_n$ clusters (corresponding to the chemical reaction Au$_n \rightarrow n$Au) at different values of the threshold $\varepsilon$ (see Fig.~\ref{fig:cohesion}). The desired accuracy of 1~kJ/mol is again achieved for $\varepsilon \approx 3 \cdot 10^{-4}$, while for larger $\varepsilon$ errors in cohesion energies are still noticeably less (up to an order of magnitude) than for correlation energies (cf. Fig.~\ref{fig:ecorr-err}a).

\section{Conclusions}
\label{sec:conc}

We presented the relativistic generalization of the single-reference rank-reduced coupled cluster method. Despite being somewhat similar to its non-relativistic counterpart~\cite{Parrish:19,Lesiuk:22,Hohenstein:22}, the relativistic version has some specific features. First, amplitudes of double excitations form a symmetric complex (and thus non-Hermitian) matrix. In this situation, singular value decomposition is recommended to obtain Tucker projectors. MP3 projectors were confirmed to be substantially superior to those obtained at the MP2 level. Furthermore, the widely used formalism employing antisymmetrized tensors of cluster amplitudes cannot be straightforwardly applied within the relativistic RR-CC method; thus, it is preferable to formulate and implement the method entirely in terms of Goldstone diagrams. This could become a serious obstacle for more sophisticated approximations than CCSD due to the dramatic increase in the number of diagrams; however, it is still expected to be manageable for the practically important CCSD(T) and CCSDT models.

The efficiency of the Tucker decomposition for compression of doubles amplitudes obtained within the relativistic CCSD method was studied in a series of benchmark calculations of molecular systems containing heavy elements, namely, (AuCl)$_n$ chains, Au$_n$ clusters, and the cluster model of solid ytterbium(II) chloride. It was shown that it is possible to achieve the 1~kJ/mol level of accuracy in both absolute and relative energies by setting the threshold $\varepsilon$ for non-zero singular values to $\sim 10^{-4}$; this threshold is quite general at least for the systems studied and allows to achieve high compression rates of amplitude tensors. This makes the relativistic RR-CCSD method very promising for high-precision modeling of large systems, such as cluster models of solids and impurity centers in crystals constructed within the frame of the compound-tunable embedding potential method. Furthermore, it would also be very useful for highly accurate calculations employing rather extended basis sets of medium-sized molecular systems containing heavy atoms; it is worth noting that typical errors due to the basis set incompleteness are significantly larger than an (actually tunable) errors introduced by discarding amplitudes with low singular values. Most molecular properties beyond energies are also, accessible by a direct differentiation of energies with respect to external fields. Still, special attention should be paid to ensure that the subspaces of compressed amplitudes are the same for different field strengths to avoid numerical instabilities.

Finally, the effective rank of an amplitude tensor $N_{\rm SVD}$ was shown to grow substantially slower with respect to the system size than the full rank $N_{occ}N_{virt}$, albeit its linear scaling was not achieved due to the technical limitations of available RCC codes. Thus achieving the sub-$N^6$ scaling for the relativistic RR-CCSD method for spatially extended objects seems plausible. The work on the program implementation of the relativistic RR-CCSD method is ongoing.

\section{Supplementary Material}
\label{sec:sm}

Supplementary material for this article contains (a) additional figures not included in the main text; (b) a table summarizing numbers of occupied and virtual spinors for the systems considered in the present study; (c) explicit factorized equations of the relativistic RR-CCD method; (d) molecular geometries and basis sets for gold clusters and gold (I) chloride; (e) definition of the cluster model of ytterbium (II) chloride (i.~e. geometry, basis sets, pseudopotentials and fractional point charges on pseudoions). Additional figures depict singular value distributions for the (AuCl)$_n$ chains and Au$_n$ clusters (Fig.~S1), effective ranks of amplitude tensors for different singular value thresholds for (AuCl)$_n$ and Au$_n$ (Fig.~S2), percentage of retained cluster amplitudes for (AuCl)$_n$ and Au$_n$ (Fig.~S3) and YbCl$_7$@CTEP (Fig.~S4), errors in cohesion energies of Au$_n$ clusters with respect to the singular value threshold (Fig.~S5).

\section{Conflicts of Interest}

The authors declare no conflicts of interest.

\begin{acknowledgments}
The authors thanks P.~A.~Khadeeva, V.~M.~Shakhova and A.~V.~Titov for fruitful discussions. A.V.O. is indebted to D.~A.~Pichugina (M. V. Lomonosov Moscow State University) for the scientific guidance during the initial stages of his scientific career and for introducing to the world of small gold clusters. Electronic structure calculations have been carried out using computing resources of the federal collective usage center Complex for Simulation and Data Processing for Mega-science Facilities at National Research Centre ``Kurchatov Institute'', \url{http://ckp.nrcki.ru/}.

The work of A.V.O. and A.S.R. at NRC ``Kurchatov Institute'' -- PNPI on the development of the program implementations of the CCSD method based on Goldstone diagrams (in both the EXP-T and CCRS codes), studies of the tensor decomposition efficiency and the analysis of the RR-CCD working equations, as well as their computational cost, was supported by the Russian Science Foundation (Grant No. 24-73-00076), \url{https://rscf.ru/project/24-73-00076/}.
\end{acknowledgments}

\appendix

\section{Reduction of the computational cost of the relativistic rank-reduced CCD equations}
\label{sec:appendix}

As it was mentioned in Section~\ref{sec:work-eqs}, the main problem of the relativistic RR-CCD and RR-CCSD methods is the presence of the terms quadratic in doubles amplitudes. Even if both the Tucker decomposition of doubles and the DF approximation are applied, their computational complexity remains $O(N^6)$ due to the intermediate tensors
\begin{equation}
O_{ki,lj} = \sum_{cd} \braket{kl|cd} t_{ij}^{cd}
\end{equation}
and
\begin{equation}
Z_{kj,bc} = \sum_{ld} \braket{kl|dc} t_{jl}^{bd}.
\end{equation}
The remedy for this problem was proposed in~\cite{Lesiuk:22}. Here we consider for example only the term involving the former intermediate, it corresponds to the diagram $D_{3a}$ in Fig.~9.2 of the book of Shavitt and Bartlett~\cite{ShavittBartlett:09}. The other terms arising in the relativistic RR-CCD approximation (double excitations only) and estimates of their scaling are listed in Supplementary Materials.

The algebraic expression corresponding to the diagram $D_{3a}$ is as follows:
\begin{equation}
    [D_{3a}]_{ij}^{ab} = \sum_{kl}  t_{kl}^{ab} \sum_{cd} \braket{kl|cd} t_{ij}^{cd}.
\end{equation}
Projecting this term onto the subspace of compressed amplitudes and taking into account the decomposition of cluster amplitudes, one arrives at:

    \begin{align}
       & [D_{3a}]_{XY} = 
        \sum_{ijkl,X''} \left( \sum_{a} {U}_{ai}^ {X*} U_{ak}^{X''} \right)\left( \sum_{b\, Y''}U_{bj}^ {Y*}  {U}_{bl}^{Y''}  T_{X''Y''} \right) \times \nonumber \\
          & \times \sum_{\substack{X'\,Q}} \left( \sum_{c} B_{kc}^{Q} U_{ci}^{X'} \right) \left( \sum_{d\, Y'} B_{ld}^{Q}  {U}_{dj}^{Y'}  T_{X'Y'} \right) = \nonumber\\
           &= \sum_{ijkl} O_{ki,lj} \sum_{X''} \left( \sum_{a} {U}_{ai}^ {X*} U_{ak}^{X''} \right) \left( \sum_{b\, Y''}U_{bj}^ {Y*}  {U}_{bl}^{Y''}  T_{X''Y''} \right) 
    \end{align}

No matter how this expression is factorized, its overall asymptotic complexity remains at least $O(N^6)$. To solve this problem, in~\cite{Lesiuk:22}, it was proposed to introduce an intermediate tensor and apply an additional Tucker decomposition to it:
\begin{align}
        O_{ki,lj} &= \sum_{\substack{X'\,Q}} \left( \sum_{c} B_{kc}^{Q} U_{ci}^{X'} \right) \left( \sum_{d\, Y'} B_{ld}^{Q}  {U}_{dj}^{Y'}  T_{X'Y'} \right) = \nonumber \\
        &= \sum^{ N_{\text{O}}}_{RW} \alpha_{ki}^{R} o^{RW}  \alpha_{lj}^{W},
\end{align}
where $N_{\rm O}$ stands for the effective tensor rank of $O_{ki,lj}$; letters $R, W, ...$ are used as auxiliary indices for the compressed tensor $o^{RW}$, range is $[1,N_{\text{O}}]$.

Projectors $\alpha_{ki}^{R}$ and the core of decomposition $o^{RW}$ are assembled only once before CC iterations started. Such a technique allows to reduce the computational cost to $O(N^5)$:
\begin{equation}
   \begin{split}
       [D_{3a}]_{XY} &= {\sum_{\substack{X''Y''}}}
        T_{X''Y''}
        {\sum_{\substack{RW}}}
        o^{RW} \times \nonumber \\
        &\times \left( {\sum_{\substack{ika\\}}}{U}_{ai}^{X*} U_{ak}^{X''} \alpha_{ki}^{R} \right)  \left( {\sum_{\substack{jlb\\}}}U_{bj*}^{Y} {U}_{bl}^{Y''} \alpha_{lj}^{W} \right).
   \end{split}
\end{equation}
Similar expressions could be derived for tensor $Z_{kj,bc}$ (Eq.~(A2); see Supplementary Materials).


\providecommand{\noopsort}[1]{}\providecommand{\singleletter}[1]{#1}%


\begin{thebibliography}{101}
\bibitem [{{T}hroughout this paper, we follow the usual practice of referring
  to multidimensional arrays with $N$ dimensions as $N$th-order tensors,
  irrespective of their transformation properties()}]{footnote:def-tensors}%
  \BibitemOpen
  {T}hroughout this paper, we follow the usual practice of referring to
  multidimensional arrays with $N$ dimensions as $N$th-order tensors,
  irrespective of their transformation properties,\ \href@noop {}
  {}\BibitemShut {NoStop}%
\bibitem [{{T}his is also true for cluster operators representing higher
  excitations, which are not the subject of the present
  work()}]{footnote:def-t2}%
  \BibitemOpen
  {T}his is also true for cluster operators representing higher excitations,
  which are not the subject of the present work,\ \href@noop {} {}\BibitemShut
  {NoStop}%
\end{thebibliography}
\end{document}